\documentclass[prd,superscriptaddress,amsmath,twocolumn,nofootinbib]{revtex4-2}

\pdfoutput=1
\usepackage{times}

\usepackage{graphicx}
\usepackage{xcolor}

\usepackage{array}
\usepackage{booktabs}
\usepackage{enumitem}
\usepackage{epsf,latexsym,bbm,euscript}

\usepackage{mathrsfs,mathtools,amsmath,amssymb}
\usepackage{tensor}

\makeatletter
\newcommand*{\defeq}{\mathrel{\rlap{%
			\raisebox{0.3ex}{$\m@th\cdot$}}%
		\raisebox{-0.3ex}{$\m@th\cdot$}}%
	=}
\newcommand*{\eqdef}{=\mathrel{\rlap{%
			\raisebox{0.3ex}{$\m@th\cdot$}}%
		\raisebox{-0.3ex}{$\m@th\cdot$}}%
}
\makeatother

\newcommand{\RNum}[1]{\uppercase\expandafter{\romannumeral #1\relax}}

\usepackage{scalerel}
\usepackage{tikz}
\usetikzlibrary{svg.path}
\definecolor{orcidlogocol}{HTML}{A6CE39}
\tikzset{
	orcidlogo/.pic={
		\fill[orcidlogocol] svg{M256,128c0,70.7-57.3,128-128,128C57.3,256,0,198.7,0,128C0,57.3,57.3,0,128,0C198.7,0,256,57.3,256,128z};
		\fill[white] svg{M86.3,186.2H70.9V79.1h15.4v48.4V186.2z}
		svg{M108.9,79.1h41.6c39.6,0,57,28.3,57,53.6c0,27.5-21.5,53.6-56.8,53.6h-41.8V79.1z M124.3,172.4h24.5c34.9,0,42.9-26.5,42.9-39.7c0-21.5-13.7-39.7-43.7-39.7h-23.7V172.4z}
		svg{M88.7,56.8c0,5.5-4.5,10.1-10.1,10.1c-5.6,0-10.1-4.6-10.1-10.1c0-5.6,4.5-10.1,10.1-10.1C84.2,46.7,88.7,51.3,88.7,56.8z};
	}
}
\newcommand\orcidlink[1]{\href{https://orcid.org/#1}{\mbox{\scalerel*{
				\begin{tikzpicture}[yscale=-1,transform shape]
					\pic{orcidlogo};
				\end{tikzpicture}
			}{X}}}}

\usepackage{url,hyperref}
\hypersetup{colorlinks,linkcolor={blue!55!black},citecolor={red!50!black},urlcolor={blue!45!black},breaklinks=true}

\begin{document}

\title{Schwarzschild black hole surrounded by a cavity and phase transition in the non-commutative gauge theory of gravity}

\author{Abdellah Touati\orcidlink{0000-0003-4478-2529}}
\email{abdellah.touati@univ-batna.dz, (touati.abph@gmail.com)}
\affiliation{Department of Physics, Faculty of Sciences of Matter, University of Batna-1, Batna 05000, Algeria}

\author{Slimane Zaim }
\email{zaim69slimane@yahoo.com}
\affiliation{Department of Physics, Faculty of Sciences of Matter, University of Batna-1, Batna 05000, Algeria}

\begin{abstract}
	In this work, we investigate the phase transition of the Schwarzschild black hole (SBH) inside an isothermal spherical cavity in the context of the non-commutative (NC) gauge theory of gravity, by using the Seiberg-Witten (SW) map and the star product. Firstly, we compute the NC correction to the Hawking temperature and derive the logarithmic correction to the entropy, then we derive the local temperature and local energy of NC SBH in isothermal cavity. Our results show that the non-commutativity removes the commutative divergence behavior of temperature, and prevents the SBH from the complete evaporation, which leads to a remnant black hole, and this geometry has predicted a minimal length in the order of Planck scale $\Theta\sim l_{planck}$. Therefore, the thermodynamic stability and phase transition is studied by analyzing the behavior of the local heat capacity and the Helmholtz free energy in the NC spacetime, where the results show that, the NC SBH has a two second-order phase transition and one first-order phase transition, with two Hawking-Page phase transition in the NC gauge theory. 
\end{abstract}

\maketitle

\section{Introduction} \label{sec:introduction}

Recently there have been a lot of interesting studies on black holes thermodynamics \cite{hawking3,harms,vaz,haranas2,hansen1,chen1,hansen,jawad1}, where it's considered the first bridge to the quantum theory of gravity. Moreover, the thermodynamic proprieties of black holes are considered as effects of the first mechanism that unifies quantum field theory and gravity in the context of a semi-classical approach, and it was discovered by Hawking \cite{hawking1,hawking2}, which demonstrated that a black hole can emit radiation as a black body, and that allowed it to evaporate.
However, this theory is not a complete theory of quantum gravity (QG), and causes a problem at the final stage of evaporation , where Hawking demonstrated that SBH has a divergence temperature behavior. This issue can be solved by a theory of quantum gravity, and that leads to the emergence of many theories that provide the effect of QG on the black holes thermodynamics, such as Rainbow Gravity \cite{rg1,rg2,lutfugr1,lutfugr2}, quantum deformation in Heisenberg uncertainty principal (as example generalized uncertainty principle (GUP), extended uncertainty principle (EUP),..etc.) \cite{lutfugup1,lutfueup3,lutfueup4,lutfueup5,lutfuegup6,gup1,gup2}, and NC geometry \cite{piero1,lopez1,nozari1,nozari2,myung1,chai2,mukhe1,zaim1,linares,Hassanabadi1,abdellah2,abdellah4},... etc. It is worth to note that some of the QG theory models (like the model above) there predict a minimal length, which is expected in the order of the Planck length, where this minimal length plays the role of natural cutoff to eliminate the divergence behavior, such as the divergence behavior of Hawking temperature for SBH. 

Recently, there are a lot of interesting studies of the thermodynamic properties of black holes inside an isothermal cavity that allows the thermodynamic stability of the black hole
\cite{cavity0,cavity1,cavity2,cavity7,feng1,cavity3,cavity4,cavity5,feng2,cavity6,feng3}, where this system has similar properties with the AdS spacetime \cite{cavity8}. 
In this Letter, we aim to study this system in the NC gauge theory, where this theory is motivated by the string theory \cite{seiberg1}, where the main idea of this theory is the quantization of spacetime itself, leads to the quantization of gravity, and that is done by imposing a commutation relation between the coordinates themselves:
\begin{equation}
	[x^{\mu},x^{\nu}]=i\Theta^{\mu\nu},
\end{equation}
where $\Theta^{\mu\nu}$ is an anti-symmetric real matrix, and we only take space-space non-commutativity ($\Theta^{0i}=0$), due to the unitary problem \cite{unitarity1,unitarity2}.
The above commutation relation modifies the ordinary product of two arbitrary functions $f(x)$ and $g(x)$ defined over this spacetime,becomes defined by a new product called the star product (Moyal product) "$*$" given by:
\begin{equation}\label{eqt2.25}
	(f*g)(x)=f(x)e^{\frac{i}{2}\Theta^{\mu\nu}\overleftarrow{\partial_{\mu}}\overrightarrow{\partial_{\nu}}}g(x).
\end{equation}

Here, we use the gauge theory of gravity, the star product and the SW map \cite{seiberg1} to investigate the stability and phase transition of SBH inside an isothermal cavity in the NC gauge theory. 

In this present work, we obtain the NC SBH in gauge theory of gravity. The corrections to the usual Hawking temperature and entropy from NC SBH are obtained. Then we investigate the thermodynamics of SBH inside an isothermal cavity. Firstly, we obtain the NC correction to the local Hawking temperature for SBH. The obtained results show that, the non-commutativity removes the divergence behavior of the commutative case, and the prediction of the NC parameter $\Theta$ shows a good agreement with the Planck scale. The NC correction to the local energy using the first law of black hole thermodynamics is also obtained. Then we study the thermal stability and phase transition of the NC SBH inside cavity by analyzing the  behaviors of local heat capacity and free energy of this system in the NC geometry. Our results show that, the NC SBH inside cavity has two physical limitation points and two phase transition of second-order, and the free energy has two Hawking-Page phase transitions with one first-order phase transition in this geometry.
 
This paper is organized as follows: In Sec.~\ref{sec:NCSBH}, we present the NC correction to the SBH in the context of gauge theory of gravity, using SW map and star product. In Sec.~\ref{sec:BTNCS}, we review the NC corrections up to the second order in $\Theta$ to both Hawking temperature and entropy of SBH. Then we consider the boundary condition of SBH inside cavity in the NC spacetime, where we obtain the NC corrections to local temperature and local energy (see sub.~\ref{subsec:NCLT}). The thermodynamic stability and phase transition are analyzed using the local heat capacity and Helmholtz free energy (see sub.~\ref{subsec:NCLCH}) that are also obtained and discussed. In Sec.~\ref{sec:concl}, we present our conclusion and remarks.


\section{Non-commutative Schwarzschild black hole}\label{sec:NCSBH}

We review briefly the formalism of the NC gauge theory of gravity \cite{cham1,chai1,chai2,mukhe1}, which is a generalization of the classical theory based of the de Sitter group as local symmetry \cite{zet1,zet2}.
Where we use the tetrad formalism and both the star $\ast-$ product and the SW map to construct a NC gauge theory for a Schwarzschild metric. According to the article \cite{cham1}.
The NC corrections to the commutative tetrad fields $\hat{e}^{a}_{\mu}(x,\Theta)$ are obtained using the perturbation form for the SW map, which is described as a development in the power of $\Theta$ up to the second-order
\begin{equation}
	\hat{e}^{a}_{\mu}(x,\Theta)=e^{a}_{\mu}(x)-i\Theta^{\nu\rho}e^{a}_{\mu\nu\rho}(x)+\Theta^{\nu\rho}\Theta^{\lambda\tau}e^{a}_{\mu\nu\rho\lambda\tau}(x)+\mathcal{O}(\Theta^{3})\label{eqt2.26}
\end{equation}
where
\begin{equation}
	e^{a}_{\mu\nu\rho}=\frac{1}{4}[\omega^{ac}_{\nu}\partial_{\rho}e^{d}_{\mu}+(\partial_{\rho}\omega^{ac}_{\mu}+R^{ac}_{\rho\mu})e^{d}_{\nu}]\eta_{cd}
\end{equation}
and

\begin{widetext}
\begin{align}
	e^{a}_{\mu\nu\rho\lambda\tau}&=\frac{\Theta^{\nu\rho}\Theta^{\lambda\tau}}{32}\left[2\{R_{\tau\nu},R_{\mu\rho}\}^{ab}e^{c}_{\lambda}-\omega^{ab}_{\lambda}(D_{\rho}R_{\tau\nu}^{cd}+\partial_{\rho}R_{\tau\nu}^{cd})e^{m}_{\nu}\eta_{dm}-\{\omega_{\nu},(D_{\rho}R_{\tau\nu}+\partial_{\rho}R_{\tau\nu})\}^{ab}e^{c}_{\lambda}+2\partial_{\nu}\omega_{\lambda}^{ab}\partial_{\rho}\partial_{\tau}e^{c}_{\mu}\right.\notag\\
	&\left.-\partial_{\tau}\{\omega_{\nu},(\partial_{\rho}\omega_{\mu}+R_{\rho\mu})\}^{ab}e^{c}_{\lambda}-\omega^{ab}_{\lambda}\left(\omega^{cd}_{\nu}\partial_{\rho}e^{m}_{\mu}+\left(\partial_{\rho}\omega_{\mu}^{cd}+R_{\rho\mu}^{cd}\right)e^{m}_{\nu}\right)\eta_{dm}-2\partial_{\rho}\left(\partial_{\tau}\omega_{\mu}^{ab}+R_{\tau\mu}^{ab}\right)\partial_{\nu}e^{c}_{\lambda}\right.\notag\\
	&\left.-\{\omega_{\nu},(\partial_{\rho}\omega_{\lambda}+R_{\rho\lambda})\}^{ab}\partial_{\tau}e^{c}_{\mu}-\left(\partial_{\tau}\omega_{\mu}+R_{\tau\mu}\right)\left(\omega^{cd}_{\nu}\partial_{\rho}e^{m}_{\lambda}+\left((\partial_{\rho}\omega_{\lambda}+R_{\rho\lambda})\right)e^{m}_{\nu}\right)\eta_{dm}\right]\eta_{cb}+\mathcal{O}\left( \Theta^{3}\right),\label{eq:SWM}
\end{align}
\end{widetext}
where $\hat{e}_{a}^{\mu }$ and $\omega^{ab}_{\mu}$ are the commutative tetrad field and spin connection, and
\begin{align}
	\{\alpha,\beta\}^{ab}&=\left(\alpha^{ac}\beta^{db}+\beta^{ac}\alpha^{db}\right)\eta_{cd},\\ [\alpha,\beta]^{ab}&=\left(\alpha^{ac}\beta^{db}-\beta^{ac}\alpha^{db}\right)\eta_{cd}\\
	D_{\mu}R_{\rho\sigma}^{ab}&=\partial_{\mu}R^{ab}_{\rho\sigma}+\left(\omega_{\mu}^{ac}R^{db}_{\rho\sigma}+\omega_{\mu}^{bc}R^{da}_{\rho\sigma}\right)
\end{align}

where the $\hat{e}_{a}^{\mu }$ is the inverse of the vierbein $\hat{e}_{\mu }^{a}$ defined as
\begin{equation}
	\hat{e}_{\mu }^{b} \hat{e}_{a}^{\mu }=\delta _{a}^{b},\quad \hat{e}_{\mu }^{a} \hat{e}_{a}^{\nu }=\delta _{\mu }^{\nu }\,.
\end{equation}

For the NC metric $\hat{g}_{\mu \nu }$, we use the expression found in Ref. \cite{chai1}
\begin{equation}\label{eq:metric}
	\hat{g}_{\mu \nu }=\frac{1}{2}(\hat{e}_{\mu }^{b}\ast \hat{e}_{\nu b}+\hat{e}_{\nu }^{b}\ast \hat{e}_{\mu b})\,.
\end{equation}
To compute the deformed metric $\hat{g}_{\mu \nu }$, we choose the following NC anti-symmetric matrix $\Theta^{\mu\nu}$
\begin{equation}
	\Theta^{\mu\nu}=\left(\begin{matrix}
		0	& 0 & 0 & 0 \\
		0	& 0 & 0 & \Theta \\
		0	& 0 & 0 & 0 \\
		0	& -\Theta & 0 & 0
	\end{matrix}
	\right), \qquad \mu,\nu=0,1,2,3\label{eqt2.34}
\end{equation}

We follow the same steps outlined in Ref. \cite{abdellah1}, we choose the following general tetrads field
\begin{align}
	\underline{e}_{\mu }^{0}& =\left(\begin{array}{cccc}\left(1-\frac{2 m}{r}\right)^{\frac{1}{2}}, & 0, & 0, & 0\end{array}\right),\notag \\
	\underline{e}_{\mu }^{1}& =\left(\begin{array}{cccc}0, & \left(1-\frac{2 m}{r}\right)^{-\frac{1}{2}}sin\theta cos\phi, & r cos\theta cos\phi, & -r sin\theta sin\phi\end{array}\right), \notag\\
	\underline{e}_{\mu }^{2}& =\left(\begin{array}{cccc}0, & \left(1-\frac{2 m}{r}\right)^{-\frac{1}{2}}sin\theta sin\phi, & r cos\theta sin\phi, & r sin\theta cos\phi\end{array}\right), \notag\\
	\underline{e}_{\mu }^{3}& =\left(\begin{array}{cccc}0, & \left(1-\frac{2 m}{r}\right)^{-\frac{1}{2}}cos\theta, & -r sin\theta, & 0\end{array}\right). \label{eq:tetrad}
\end{align}
The components of the deformed metric $\hat{g}_{\mu \nu }$ for the Schwarzschild black hole can be computed using Eqs. \eqref{eq:SWM} together with Eq. \eqref{eq:tetrad}, and we use of the definition \eqref{eq:metric}, we obtain the non-zero components of the NC metric $\hat{g}_{\mu \nu }$ in the leading order on $\Theta$

\begin{widetext}

	\begin{align}
		-\hat{g}_{00}&=\left(1-\frac{2 m}{r}\right)+\Theta^{2}\left\{\frac{m\left(88m^2+mr\left(-77+15\sqrt{1-\frac{2m}{r}}\right)-8r^2\left(-2+\sqrt{1-\frac{2m}{r}}\right)\right)}{16r^4(-2m+r)}\right\}\sin^2\theta+\mathcal{O}(\Theta^4),\label{eq:13}\\
		\hat{g}_{11}&=\left(1-\frac{2 m}{r}\right)^{-1}+\Theta^{2}\left\{\frac{m\left(12m^2+mr\left(-14+\sqrt{1-\frac{2m}{r}}\right)-r^2\left(5+\sqrt{1-\frac{2m}{r}}\right)\right)}{8r^2(-2m+r)^3}\right\}\sin^2\theta+\mathcal{O}(\Theta^{4}),\label{eq:14}	
	\end{align}
	\begin{align}
		\hat{g}_{12}&=-\Theta^{2}\left\{\frac{m \left(-4m^2+r^2\left(\sqrt{1-\frac{2 m}{r}}-7\right)+m r \left(16-17 \sqrt{1-\frac{2 m}{r}}\right)\right)}{32 r (2 m-r)^3}\right\}\sin(2\theta)+\mathcal{O}(\Theta^4),	\label{eq:14'}\\
		\hat{g}_{22}&=r^{2}+\Theta^2\left\{\frac{-8 m^3-6 m^2 r \sqrt{1-\frac{2 m}{r}}+50 m^2 r-6 r^3 \sqrt{1-\frac{2m}{r}}+23 m r^2 \sqrt{1-\frac{2 m}{r}}-43 m r^2+10r^3}{32 r (r-2 m)^2}\right.\notag\\
		&\left.+\frac{\cos(2\theta)\left(8m^3+6m^2r\left(\sqrt{1-\frac{2m}{r}}+5\right)+2r^3\left(5-3\sqrt{1-\frac{2m}{r}}\right)+mr^2\left(13\sqrt{1-\frac{2 m}{r}}-37\right)\right)}{32r(r-2m)^2}\right\}+\mathcal{O}(\Theta^4),\label{eq:15}\\
		\hat{g}_{33}&=r^{2}\sin^2\theta+\Theta^{2}\left\{\frac{2m^2r\left(74-9
		\sqrt{1-\frac{2m}{r}}\right)+4r^3\left(5-3\sqrt{1-\frac{2m}{r}}\right)+mr^2\left(47\sqrt{1-\frac{2m}{r}}-97\right)-68m^3}{32r(r-2m)^2}\notag\right.\\
		&\left.+\frac{m\cos(2\theta)\left(68 m^2+r^2\left(17-11\sqrt{1-\frac{2 m}{r}}\right)+2mr\left(9 \sqrt{1-\frac{2m}{r}}-34\right)\right)}{32r(r-2m)^2}\right\}\sin^2\theta+\mathcal{O}(\Theta^4).\label{eq:16}
\end{align}
\end{widetext}
It is clear that, for $\Theta\rightarrow0$, we obtain the commutative Schwarzschild solution. The NC line element according to the above metric is given by:
\small
\begin{equation}\label{eq:line-element}
	d\hat{s}^{2}=-\hat{g}_{00}dt^{2}+\hat{g}_{11}dr^{2}+2\hat{g}_{12}drd\theta+\hat{g}_{22}d\theta^{2}+\hat{g}_{33} d\phi^{2}\,.
\end{equation}
\normalsize
For such a black hole, we find that the event horizon in non-commutative space-time where the NC metric \eqref{eq:13} satisfies this conditions $\frac{1}{\hat{g}_{11}}=0$, and its solution gives us the NC event horizon of the Schwarzschild black hole (SBH) \cite{abdellah2}:
\begin{equation}\label{eq:horizon}
	r_{h}^{NC}=r_{h}\left[1+\frac{3}{8}\left(\frac{\Theta}{r_h}\right)^2\,sin^2\theta\right].
\end{equation}
where $r_{h}=2m$ is the event horizon in the commutative case when $\Theta =0$. The effect of non-commutativity is small, which is reasonable to expect since at large distances it can be considered neglected.


\section{Black hole thermodynamics in NC spacetime} \label{sec:BTNCS}

In this section we investigate the thermodynamic and phase transition of NC SBH inside a spherical cavity with radius $R$. In our previous work \cite{abdellah2} we study the black hole thermodynamic proprieties and phase transition in the NC gauge theory of gravity, where we used the classical approaches such as the surface gravity to describe Hawking temperature, the area laws for the entropy in the NC spacetime. According to the deformed metric components \eqref{eq:13} and \eqref{eq:14}, the NC Hawking temperature can be obtained using the NC surface gravity $\hat{\kappa}$ as in \cite{abdellah2}:
\begin{align}\label{eq:temperature1}
	\hat{T}_H=\frac{\hat{\kappa}}{2\pi}&=-\frac{1}{4\pi\sqrt{-\hat{g}_{00}(r,\Theta)*\hat{g}_{11}(r,\Theta)}}\left.\frac{\partial \hat{g}_{00}}{\partial r}\right|_{r=r_h^{\mathrm{NC}}},\notag\\
	&=T_H\left(1-\frac{3\Theta^2}{2r_h^2}sin^2\theta\right).
\end{align}
where $T_H=\frac{1}{4\pi r_h}$ is the commutative Hawking temperature.
The temperature emitted from the surface horizon of the SBH in NC geometry is obtained as follows \cite{abdellah2}
\begin{align}\label{eq:4.4'}
	\hat{T} &=\frac{1}{\int_{0}^{2\pi }\int_{0}^{\pi } \sqrt{\hat{g}_{22}\ast \hat{g}_{33}}d\theta d\varphi }\int_{0}^{2\pi }\int_{0}^{\pi }  \hat{T}_H\sqrt{\hat{g}_{22}\ast \hat{g}_{33}}d\theta d\varphi \notag\\
	&=T_H\left(1-\frac{\Theta^2}{r_h^2}\right).
\end{align}

Moreover, the NC correction to the entropy is obtained using the area law of the NC SBH, and this one is not respected the first law of black hole thermodynamics which is modified by an extra factor as it shows in \cite{abdellah2}, to fix this issue we use the first law of black hole thermodynamics to get the NC correction to the entropy \cite{abdellah4}
\begin{align}\label{eq:entropy1}
	\hat{S}_{BH}=\int \frac{d\hat{m}}{\hat{T}}=\pi r_h^2+\frac{3\pi\Theta^2}{4}log(\pi r_h^2).
\end{align}
where $\hat{m}=(\frac{r_h}{2}+\frac{\Theta^2}{8r_h})$ is the NC mass of the NC SBH \cite{abdellah2}, for more detail on the logarithmic correction to the entropy in NC gauge theory see Ref. \cite{abdellah4}. As we see the above equation show a logarithmic correction to the entropy in the NC spacetime, and the area law of entropy $S=\pi r_h^2$ is recovered when we set $\Theta=0$.


\subsection{Local temperature and energy}\label{subsec:NCLT}

In the purpose to maintain thermal stability of the black hole, we need to use the outer boundary condition \cite{cavity0,cavity1,cavity2}, in which we consider the NC SBH inside a spherical and isothermal cavity with radius $R$. The local temperature of the NC SBH at a ﬁnite distance $R$ outside the black hole can be obtained as follows \cite{cavity0}:

\begin{equation}\label{eq:tempertature2'}
	\hat{T}_{local}=\frac{\hat{T}_H}{\sqrt{\hat{g}_{00}(R,\Theta)}},
\end{equation}
using the component \eqref{eq:13} at the leading order in $m$, we can find the local temperature of the NC SBH:
\begin{widetext}
\begin{align}\label{eq:tempertature2}
	\hat{T}_{local}=\frac{1}{4r_h\pi\sqrt{1-\frac{r_h}{R}}}+&\left(\frac{(-128 r_h^2 R^3 + 256 r_h R^4 - 128 R^5)-44 r_h^5 + r_h^4 \left(77 - 15 \sqrt{1-\frac{r_h}{R}}\right) R}{512r_h^3\pi R^3(R-r_h)^2\sqrt{1-\frac{r_h}{R}}}\right.\notag\\
	&\left.+\frac{16 r_h^3 \left(-2+\sqrt{1-\frac{r_h}{R}}\right) R^2}{512r_h^3\pi R^3(R-r_h)^2\sqrt{1-\frac{r_h}{R}}}\right)\Theta^2.
\end{align}
\end{widetext}
It is clear that, the local temperature \eqref{eq:tempertature2} can be reduced to the NC Hawking temperature \eqref{eq:temperature1} when $R\rightarrow \infty$, and the commutative case is recovered \cite{cavity1} when we set $\Theta=0$.
\begin{figure}[h]
	\centering
	\includegraphics[width=0.5\textwidth]{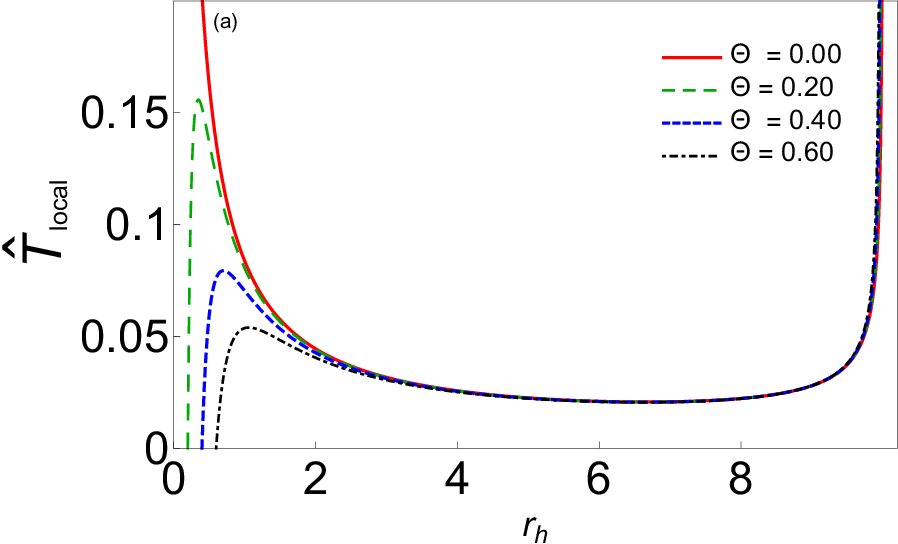}
	\includegraphics[width=0.5\textwidth]{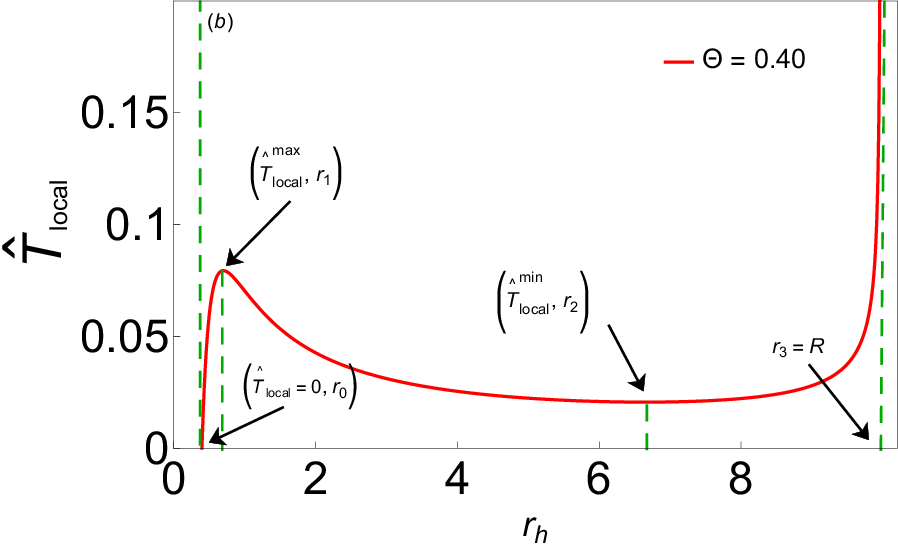}
	\caption{The behavior of local temperature as a function of the black hole event horizon $r_h$ inside a cavity with radius $R=10$.}
	\label{fig1}
\end{figure}

In Fig. \ref{fig1} we show the behavior of local temperature in the NC gauge theory of gravity as a function of the black hole event horizon $r_h$ and for a fixed cavity radius $R=10$. It is clear that, in the NC spacetime the local temperature has two extremum one minimum $(\hat{T}_{local}^{min},r_2)$ and a maximum $(\hat{T}_{local}^{max},r_1)$ (see Fig. \ref{fig1} (b)), in which the non-commutativity removes the divergence of the commutative case, and leads the black hole to reach a new maximum $\hat{T}^{max}_H\approx\frac{0.03117}{\Theta}$ at the size of $r_h^{crit}\approx \Theta$ before turning to zero at $r_0=\Theta$ at this point the black hole stops the radiation, which is contrary to the commutative case (see (a) Fig. \ref{fig1}), and the effect of the non-commutativity is analyzed in detail for the case $R\rightarrow\infty$ in our previous work \cite{abdellah2}. As it is shown in Fig. \ref{fig1} (b), the local temperature at the surface of the cavity $r_3=R$ diverges as in the commutative case, due to the boundary condition of the isothermal cavity, and during the evaporation its temperature decreases until it reach the minimum $\hat{T}_{local}^{min}$ at $r_h=r_2$, and it decreases with the increase in the NC parameter $\Theta$, then this behavior changes by increasing during the evaporation to reach the maximum local temperature $\hat{T}_{local}^{max}$ at $r_1$, then quickly falls to zero at $r_0$ and the black hole stops radiating which means a remnant black hole with minimum size $r_0$ is produced. It is worth to note that, the change of behavior at the point $r_2$ from decreasing to increasing and and the same observed at $r_1$ which changes from increasing to decreasing, where that indicates the NC SBH black hole has two phase phase transition at these points $r_1$ and $r_2$.

Furthermore, at the back-reaction point the maximum of temperature equals to the thermal energy, and it's given by the relation\footnote{In the natural system of unity ($\hbar=k_B=c=1$).} $E_{th}=\hat{T}^{max}_H \approx \frac{0.03117}{\Theta}$ at the point $r_h^{crit}=2m \approx \Theta$, and the total mass of the black hole\footnote{In this study we use the reduced Planck mass $M_{Planck}=2.435\times10^{18}Gev$ } $M = \frac{1}{2}r_h^{crit}\frac{1}{G}\approx 0.5\,\Theta\, M_{Planck}^2$. The NC parameter can be estimated as follows
\begin{equation} \label{eq:32}
	\Theta \approx 1.948\times 10^{-35}m\sim l_{Planck}.
\end{equation}

This result is in the Planck scale, which is good according to the one obtained using the gravitational wave experimental data \cite{ThetaGW} and our previous work using the thermodynamics process of the NC SBH \cite{abdellah2}. However, there are some papers that obtained a bound on the NC parameter through the study of the black holes thermodynamics, e.g, \cite{piero1,nicolini2,alavi,bound1}, and it is expected to be $\sqrt{\theta}\sim 10^{-1}.l_p$. According to this result and our previous works \cite{abdellah1,abdellah2,abdellah3}, with some references in literature as \cite{piero1,nicolini2,alavi,bound1,ThetaGW}, confirms that the NC property of spacetime appears close to the Planck scale.

Next, the local energy of the NC SBH it can be obtained using the first law of black hole thermodynamics as follow \cite{cavity7}
\begin{equation}\label{eq:energy}
	\hat{E}_{local}=\int_{r_0}^{r_h}\hat{T}_{local}d\hat{S}_{BH},
\end{equation}
where $r_0=m_0=\Theta/2$ denotes the minimum mass of the NC SBH \cite{abdellah2}. Using the NC local temperature \eqref{eq:tempertature2} and the NC entropy \eqref{eq:entropy1}, the above expression can be computed as fellows:
\begin{widetext}
\begin{align}\label{eq:energy2}
	\hat{E}_{local}&=-R\sqrt{1-\frac{r_h}{R}}+\Theta^2\left(\frac{
		42r_h\sqrt{1-\frac{r_h}{R}}(r_h-R) R^{2} Log(1-\frac{r_h}{R}) - 
	48 (|r_h-R|)^{3/2}(R)^{3/2} ArcTan(\sqrt{\frac{|r_h- R|}{R}})}{384 R^{5/2}}\right.\notag\\
	&\left.+\frac{88 r_h^4 + 66 r_h^3 R+45 r_h^3 \sqrt{1-\frac{r_h}{R}} R - 504 r_h^2 R^2 - 42 r_h^2 \sqrt{1-\frac{r_h}{R}} R^2 + 
	400 r_h R^3 - 48 R^4}{768 r_h \sqrt{1-\frac{r_h}{R}} R^3 (R-r_h)}\right)-(r_h \rightarrow r_0).
\end{align}
\end{widetext}
where the commutative expression of the local energy for the SBH $E_{local}=R\left(1-\sqrt{1-\frac{r_h}{R}}\right)$ is recovered when $\Theta=0$.

\subsection{Local heat capacity and phase transition}\label{subsec:NCLCH}
For the analyzes of the phase transition and the thermal stability of the black hole, it is necessary to check the behavior of its heat capacity, as in our case we study the thermal stability of NC SBH inside a spherical cavity, in this context we use the NC local quantity in the first law of black hole thermodynamics \eqref{eq:energy}. The NC local heat capacity can be evaluated as follows: \cite{cavity0}

\begin{equation}\label{eq:heatcapacity1}
   	\hat{C}_{local}=\frac{\partial \hat{E}_{local}}{\partial \hat{T}_{local}}=\hat{T}_{local}\frac{\partial\hat{S}}{\partial\hat{T}_{local}},
\end{equation}
using the expression of entropy \eqref{eq:entropy1} and local temperature \eqref{eq:tempertature2} for NC SBH, one get the following behavior
\begin{figure}[h]
	\centering
	\includegraphics[width=0.48\textwidth]{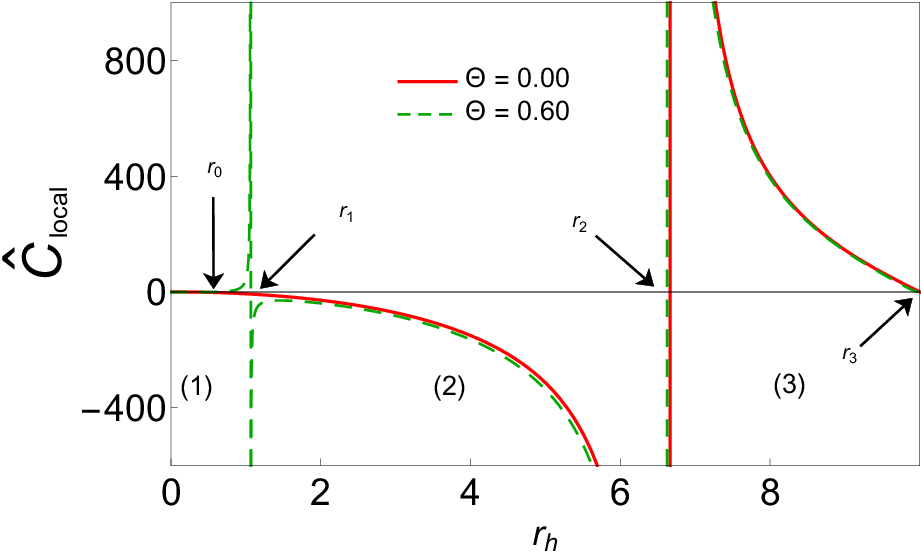}
	\caption{The behavior of local heat capacity as a function of the black hole event horizon $r_h$ inside a cavity with a radius $R=10$.}
	\label{fig2}
\end{figure}

In Fig. \ref{fig2} we show the behavior of the local heat capacity of the NC SBH inside a cavity as a function of the black hole size $r_h$. We observe that for $\Theta=0$ we recover the standard local heat capacity of SBH inside cavity, which has two branches, and that means one phase transition in the commutative SBH. However, in the NC case $\Theta\neq0$, the local heat capacity has three branches, which means the NC SBH inside a cavity has two phase transitions as we mentioned in Subsec. \ref{subsec:NCLT}. Note that, our result is in agreement with the other models of quantum gravity \cite{feng1,cavity3,feng2,feng3,chen}

It can be observed that the local heat capacity is equal to zero at $r_h=r_0$ (it's related to the non-commutativity effect $r_0=\Theta$) and $r_h=r_3$ (is related to the boundary condition $r_3=R$), and that means this black hole has two physical limitation points \cite{saheb1}, another important point is the divergence of the local heat capacity at $r_h=r_1$ and $r_2$ (which correspond to the maximum and the minimum temperature of this black hole). This implies two phase transition at these two points.

In NC geometry, $\hat{C}_{local}$ has three branches, in which the larger and smaller black hole (L-BH and S-BH respectively) have a positive heat capacity for $r_2<r_h<r_3$ and $r_0<r_h<r_1$, which correspond to the equilibrium (stable) system, where these two regions are separated by a new unstable intermediate (I-BH) region with negative heat capacity $r_1<r_h<r_2$ (see Table. \ref{tab1}). The divergence of heat capacity describes a phase transition of SBH in the NC spacetime at critical points $r_h=r_2$ and $r_1$, where this two points getting closer to each other with the increasing in $\Theta$, and the stable stage $r_0<r_h<r_1$ with positive heat capacity increasing with $\Theta$, which means that the S-SBH takes longer to stop radiating and evaporating.
\begin{table}[h]
	\begin{center}
		\caption{Region, heat capacity, state and stability of the
			NC SBH surrounded by a cavity in NC/Commutative geometry for different branches.}\label{tab1}
		\begin{tabular}{c c c c c c  }
			\hline
			\hline
		Geometry & Branches	& Region & Heat capacity & State & Stability \\
			\hline
			\hline
$\Theta=0$	 & 1	& $r_h>r_1'$ & $C>0$ & L-BH & stable  \\
		     & 2	& $r_h<r_1'$ & $C<0$ & S-BH & unstable  \\
			\hline
$\Theta\neq0$& 1	&  $r_2<r_h<r_3$ & $\hat{C}>0$ & L-BH & stable	\\
			
			 & 2	& $r_1<r_h<r_2$ & $\hat{C}<0$ & I-BH & unstable  \\
			
			 & 3	& $r_0<r_h<r_1$ & $\hat{C}>0$ & S-BH & stable  \\
			\hline
		\end{tabular}
	\end{center}
\end{table}

Table. \ref{tab1} shows the stability and the state of each region for the NC SBH inside cavity. It can be observed that, in the commutative case the local heat capacity has only two branches and one physical limitation point $r_3$. Obviously, these two branches are with different states and stability which means one phase transition point $r_2$, where the L-BH is thermodynamically stable, and that means it takes longer to evaporate, while the S-BH is unstable and evaporates quickly. In the NC spacetime, a new unstable region intermediate (I-BH) between two stable ones, where the L-BH and S-BH are stable thermodynamically in this geometry, which means they take a longer to evaporate compared to the I-BH which decays quickly to L-BH or S-BH.

It is necessary to investigate the Helmholtz free energy of NC black hole in this system for more detail analysis of the stability and phase transition of the NC SBH inside a cavity. In the system of black hole surrounded by cavity the on-shell free energy is given by the local energy and local temperature \cite{cavity7}
\begin{equation}\label{eq:freeenergy}
 	\hat{F}_{on}=\hat{E}_{local}-\hat{T}_{local}\hat{S}.
\end{equation}

Using the NC entropy \eqref{eq:entropy1}, NC local temperature \eqref{eq:tempertature2} and the NC local energy \eqref{eq:energy2} of the SBH we can illustrate the behavior of the black hole free energy inside a cavity with fixed radius $R$.

\begin{figure*}[]
	\centering
	\includegraphics[width=0.48\textwidth]{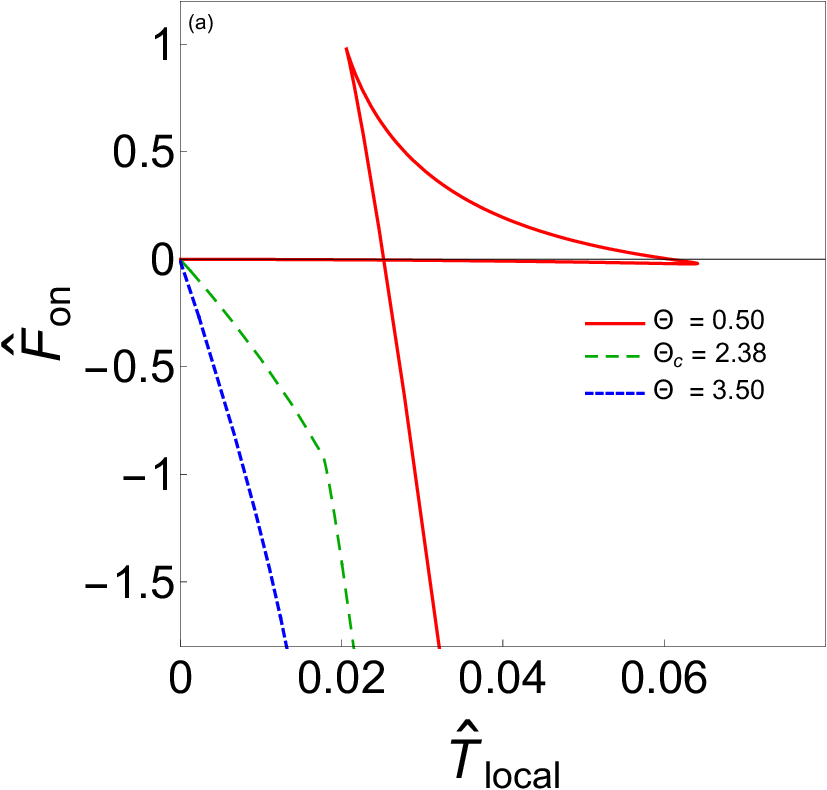}\hfill
	\includegraphics[width=0.48\textwidth]{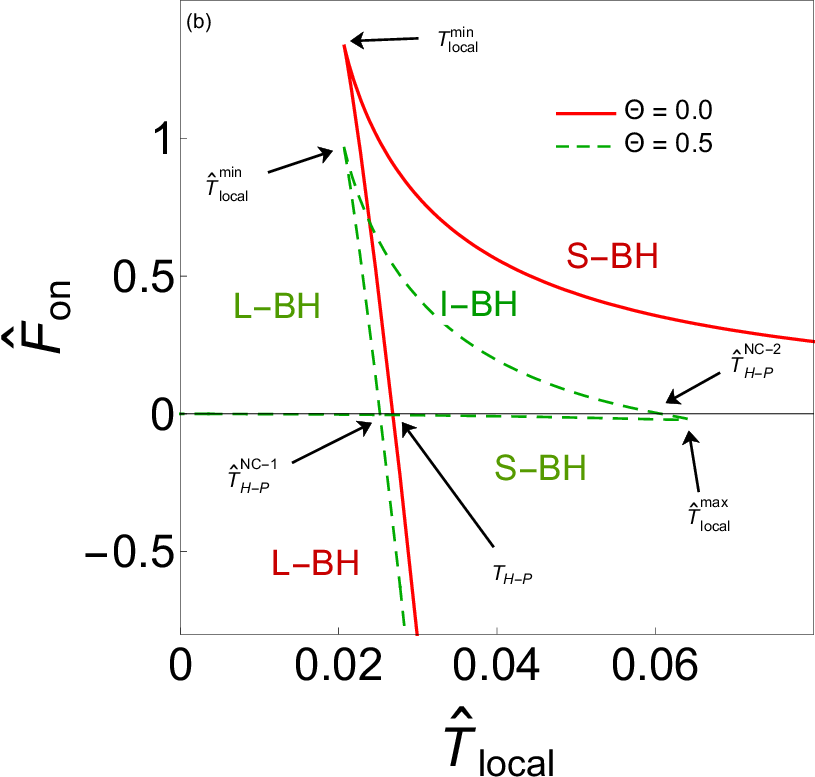}
	\caption{The behavior of Helmholtz free energy as a function of the black hole event horizon $r_h$ inside a cavity with radius $R=10$.}
	\label{fig3}
\end{figure*}

In Fig. \ref{fig3} we show the behavior of the local Helmholtz free energy $\hat{F}_{on}$ as a function of the local temperature $\hat{T}_{local}$ of the NC SBH inside cavity with radius $R=10$ in the NC spacetime with different $\Theta$. Obviously, in the NC spacetime the phase structure of the NC SBH inside cavity shows an exact swallowtail structure for $\Theta<\Theta_c$, and that indicates a two-phase coexistence state, while for $\Theta\geq \Theta_c$ the swallowtail structure disappears see Fig. \ref{fig3} (a). It is worth to note that, due to the perturbative nature of this theory, the NC parameter in natural system is considered small $\Theta<1$, so in this case the only possibility is for $\Theta<\Theta_c$, and that means in this geometry the NC SBH has only the possibility of two-phase coexistence state phase transition, which are consistent with the profile of $\hat{T}_{local}-r_h$ (Fig. \ref{fig1}) and $\hat{C}_{local}-r_h$ (Fig. \ref{fig2}). In the commutative spacetime time, the phase structure of SBH inside cavity shows a Hawking-Page (H-P) phase transition (see Fig. \ref{fig3} (b)), which occurs at $T_{H-P}$, for more detail on the H-P phase transition see Ref. \cite{cavity9}. Obviously, the stability of the black hole depends on its free energy, in which the stable black hole has less free energy and the unstable one has a high free energy. 
However, in the NC spacetime the above phase structure becomes a swallow tail structure, white two inflection points at $\hat{T}_{local}^{max}$ and $\hat{T}_{local}^{min}$ (see Fig. \ref{fig3} (b)), and two H-P phase transition point at the root of free energy $\hat{F}_{on}=0$, at $(\hat{T}^{NC-1}_{H-P}, \hat{T}^{NC-2}_{H-P})$, and that indicates a two-phase coexistence state. As we observe in Fig. \ref{fig3} (b), the location of the S-BH and the L-BH are respectively ($0,\hat{T}^{min}_{local}$) and ($\hat{T}_{local}>\hat{T}_{local}^{max}$), while the three branches of the NC SBH are located in ($\hat{T}_{local}^{min},\hat{T}_{local}^{max}$), which is similar to the one obtained using GUP \cite{chen}.


\section{Conclusions}\label{sec:concl}

Based on the NC gauge gravity, we construct the deformed SBH using a general form of tetrad fields. Then, we extend our previous work \cite{abdellah2} to the case of NC SBH surrounded by an isothermal spherical cavity. Firstly, we obtain the NC correction of local Hawking temperature. The analysis of the NC local temperature shows the effect of non-commutativity, in which this geometry removes the commutative divergence and predicts a remnant black hole in the final stage of evaporation. Moreover, this geometry leads to the SBH to reach two extremum temperatures, one minimum and the other it's a maximum, which means two phase transition (see Fig. \ref{fig1}). Then we show the estimation of $\Theta$ it is in the order of the Planck scale. The NC correction to the local energy is also obtained using the first law of black hole thermodynamics. 

Secondly, we check the stability and the phase transition of the NC SBH inside the cavity. The analysis of the local heat capacity shows that, in the NC spacetime the heat capacity has two physical limitation points at $r_0$ and $r_3$, and three branches with two divergence points at $r_1$ and $r_2$ which represent two phase transition of second order. In the NC geometry, new unstable intermediate black hole appears (see Fig. \ref{fig2}) which is not allowed in the commutative case, interpolates between two stables black holes the smaller and the larger ones. 

Finally, the analysis of the on-shell free energy shows a swallowtail structure in the NC spacetime, which does not appear in the commutative case, and that indicates the NC SBH has a first order phase transition. As we see in Fig. \ref{fig3}, we have found one H-P critical point in commutative spacetime at $T_{H-P}$, while for the NC spacetime has two H-P critical points at $\hat{T}_{H-P}^{NC-1}$ and $\hat{T}_{H-P}^{NC-2}$, and this result is in agreement with the one obtained by RG in extended phase space \cite{feng3}, whereas one and three H–P critical points were observed in other models of quantum gravity such as \cite{feng1,cavity5,feng2}. In the case of temperature below the first H–P critical points ($0<\hat{T}^{NC-1}_{H-P}$), the free energy of L-BH and I-BH are greater than the S-BH ($\hat{F}_{on}^{I}>\hat{F}_{on}^{L}>\hat{F}_{on}^{S}$) and that indicates the unstable I-BH and stable L-BH decay into a stable S-BH, while for the case between the two H–P critical points ($\hat{T}^{NC-1}_{H-P}<\hat{T}_{local}<\hat{T}^{NC-2}_{H-P}$), the stable S-BH and the unstable I-BH decay into stable L-BH, because the free energy of L-BH is lower than I-BH and S-BH ($\hat{F}_{on}^{I}>\hat{F}_{on}^{S}>\hat{F}_{on}^{L}$), and it's the same case for the temperature above the second H-P critical point $\hat{T}^{NC-2}_{H-P}<\hat{T}_{local}<\hat{T}^{max}_{local}$. As it is observed in the Fig. \ref{fig2} and Fig. \ref{fig3}, the unstable intermediate black hole (I-BH) decays quickly to the stable ones (S-BH or L-BH), and that means I-BH cannot survive for a long time. The NC gauge theory of gravity shows a good agreement with some models of QG as \cite{feng1,cavity3,feng2,chen,cavity5,feng3}, and that indicates the impact of the NC gauge theory to describe the quantum effect of gravity in a good way, with the prediction of the minimal length at Planck scale $\Theta$.

\acknowledgments
This work is supported by PRFU Research Project B00L02UN050120230003, Univ. Batna 1, Algeria.

\bibliography{ref}

\end{document}